\documentclass[prb,amssymb,twocolumn,showpacs]{revtex4}

\usepackage{graphicx}
\usepackage{bm}

\begin{document}

\title{Magnetic Properties of the Intermediate State in Small Type-I Superconductors}

\author{Alexander D. Hern\'{a}ndez}
\altaffiliation{Present addresss: The Abdus Salam International Centre for Theoretical Physics,
Strada Costiera 11, (34014) Trieste, Italy}
\affiliation{Centro At{\'{o}}mico Bariloche and Instituto Balseiro,
8400 San Carlos de Bariloche, R\'{\i}o Negro, Argentina.}
\author{Daniel Dom\'{\i}nguez}
\affiliation{Centro At{\'{o}}mico Bariloche and Instituto Balseiro,
8400 San Carlos de Bariloche,
R\'{\i}o Negro, Argentina.}

\begin{abstract}

 We present simulations of the intermediate state of type-I
superconducting films solving the time dependent Ginzburg-Landau
equations, which include the demagnetizing fields via the
Biot-Savart law. For small  square samples we find that, when
slowly increasing the applied magnetic field $H_a$,  there is a saw-tooth
behavior of the magnetization and very geometric patterns, due to
the influence of surface barriers; while when slowly decreasing
$H_a$, there is a positive magnetization and symmetry-breaking
structures. When random initial conditions are considered, we
obtain  droplet and laberynthine striped patterns, depending
on $H_a$.

\end{abstract}


\maketitle
 
In 1937 Landau modeled
the intermediate state (IS) in
thin slabs  of type-I superconductors,  
assuming  a periodic structure of
alternating stripes of normal and superconducting phases.\cite{Landau}
Direct experimental observation of the IS
revealed that, while in some cases its structure
resembled the Landau picture, very complex patterns and
history dependence were usually seen.
\cite{Huebener,Huebener72,kirchner,miller_cody} 
Similar type of complex structures were later observed
in two-dimensional (2D) systems where there is a
competition among interfacial tension and 
long-range interactions \cite{seul_andelman} 
like thin magnetic films, 
ferromagnetic fluids, 
 Langmuir and lipid monoloyers, 
and self-assembled atoms on solid surfaces.\cite{seul_andelman}  
Labyrinthine patterns, and a transition from structures 
of droplets to stripes are typically observed.\cite{seul_andelman} 
The rich physics found in these 2D systems has motivated
a renewed interest in the study of the IS in type-I superconductors
in several recent experiments. \cite{castro,reisin_lipson,egorov,jeudy,mariela}

Most of the theoretical progress 
\cite{maki_lasher_callaway,hubert,Goren_Tinkham,fortini_barrier,SL,Dorsey91,dorsey96,narayan,choksi}
has been made by modeling the IS with
periodic arrays of normal and superconducting phases. 
Recently, a current-loop model \cite{dorsey96} which allows to describe simple 
non periodic patterns has been introduced,
but a fully consistent 
theoretical description\cite{narayan} of the experimental patterns 
is still needed. 
Another important problem not addressed neither experimentally nor theoretically in
type-I superconductors is the study of the IS in samples of sizes comparable with the 
expected periodicity of the patterns, while interesting ``mesoscopic'' behaviors have been found 
in type-II superconductors with sizes of the order of few times the magnetic 
size ($\lambda$) of vortices.\cite{geim}

In this paper we report detailed simulations of small square type-I 
superconductors by solving the time dependent Ginzburg-Landau
(TDGL) equations for slabs of thickness $d$. We 
consider the ``non-branching case", where $d\ll d_s\approx
800(\xi-\lambda)$,\cite{hubert} which can be well approximated by
reducing the equations to a 2D problem, as done
for example in Ref.\onlinecite{maki_lasher_callaway,dorsey96}. We
therefore assume that the current density ${\bf J}$ and the order
parameter $\Psi$ can be replaced by their average over $z$ for
$-d/2 < z < d/2$, i.e. ${\bf J}({\bf R},z)\rightarrow{\bf J}({\bf R})$
and ${\Psi}({\bf R},z)\rightarrow{\Psi}({\bf R})$, with ${\bf
R}=(x,y)$ the in-plane coordinate. The TDGL equations, \cite{Dorsey91,dorsey96}
 in the gauge where the electrostatic potential is zero, are 
\begin{eqnarray}
\frac{\partial\Psi}{\partial t}  &=&
\frac{1}{\tau_G}
\left(|\Psi|^2-1\right)\Psi - \frac{\xi^2}{\tau_G}\left(\nabla - i\frac{2e}{\hbar c}
{\bf A}\right)^2\Psi\\
{\mathbf J} &=&\frac{1}{8\pi e\lambda^2}\mbox{Im}\left[\Psi^*
\left(\nabla - i\frac{2e}{\hbar c}{\mathbf A}\right)\Psi\right]
 -\frac{\sigma_n}{c}\frac{\partial
{\mathbf A}}{\partial t}
\end{eqnarray}
Here  $\nabla$, ${\mathbf A}$, ${\bf J}$ 
are 2D in-plane vectors, $\tau_G=\xi^2/D$, 
$D$ is the diffusion constant, $\sigma_n$ the normal state
conductivity, $\lambda(T)$ the penetration depth and $\xi(T)$ the
coherence length. These 2D  approximated TDGL equations
couple  with the perpendicular component of ${\bf B}$.
Following Ref.\onlinecite{brandt} 
we express the $z$-averaged sheet current ${\bf J}$ by a scalar
function $g$:
 ${\bf J}({\bf R})=
 \nabla\times\hat{\bf z} g({\bf R})$.
This guarantees that $\nabla\cdot{\bf J}=0$, the physical meaning of
$g({\bf R})$ being the local magnetization or density of tiny
current loops. Next one relates $g({\bf R})$ with 
$B_z({\bf r})=\nabla\times {\bf A}|_z$ at
$z=0$ and the applied field $H_a$ by means of the Biot-Savart law:\cite{brandt}
\begin{equation}
B_z({\bf R},z=0)=H_a+\frac{1}{c}\int Q({\bf R},{\bf R'})g({\bf R'}) d^2{\bf
R'}
\end{equation}
 The kernel $Q$ satisfies $Q({\bf R}_1,{\bf R}_2)\equiv Q({\bf R}_2-{\bf
 R}_1)$;
 $Q({R \gg d})=-d/R^3$; and $\int
d^2{\bf R} Q({\bf R})=0$ (due to flux conservation). To a good
approximation the kernel can be given by 
(see K. Maki in Ref.~\onlinecite{maki_lasher_callaway}):
$Q({\bf R})=4\pi\delta({\bf R})- d/[|{\bf R}|^2+ d^2/4]^{3/2}$. 
The boundary conditions are $(\nabla -i\frac{2e}{\hbar c}{\bf A})|_\perp \Psi =0$
and $g|_b=0$.
We solve the equations with a
finite difference scheme with discretization $\Delta x = \Delta y
= 0.5\xi(0)$, using link variables to maintain the gauge invariance.\cite{Dorsey91}
We normalize time  by $t_0=4\pi \sigma_n\lambda^2/c^2$,
${\mathbf A}$  by $H_{c2}(0) \xi (0)$ , $T$
(temperature) 
by $T_c$, and take
$\tau_G/t_0=12$. 
To obtain $g({R})$ we invert (3) using the conjugate gradient method
as done in Ref.\onlinecite{dj96}. 

We show results for square samples of size $L \times L$ with $L=256\xi(0)$, 
thickness $d=40\xi(0)$; $\kappa=\lambda/\xi=0.6$; $T=0.8 T_c$ and time step $\Delta t=0.25$   
(we also obtained similar results for $L=120\xi(0)-256\xi(0)$, $d=10\xi(0)-40\xi(0)$ and 
$\kappa=0.4-0.6$).
\begin{figure}[ht]
\begin{center}
\includegraphics[width=0.9\linewidth]{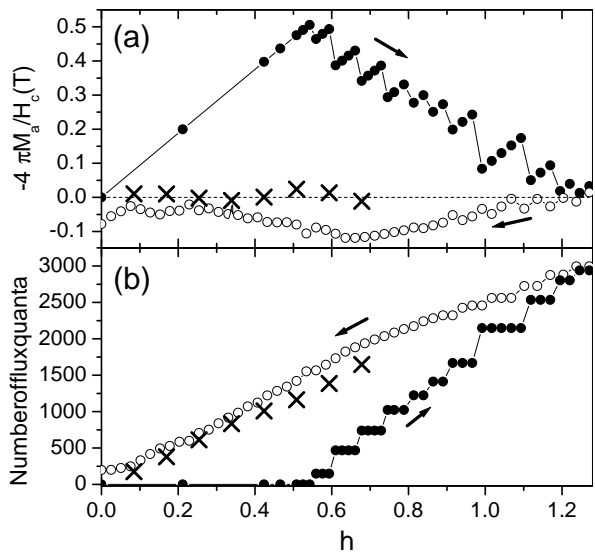}
\caption{(a) $-(B-H_a)/H_c(T)$ and (b) the number of flux quanta
obtained increasing (closed circles), decreasing (open circles)
the external magnetic field $h=H_a/H_c(T)$, and with random
initial conditions (crosses).} \label{fig1}
\end{center}\end{figure}
\noindent
We performed simulations of the intermediate state following
three different procedures: (i) slowly increasing the magnetic field
from $H_a=0$, (ii) slowly decreasing the magnetic field from the
normal state ($H_a > H_c$) and (iii) starting from random initial
conditions for each value of $H_a$. The global results for the
three cases are summarized in Fig.~1. We show the apparent
magnetization, $4\pi M_a=\langle B_z\rangle-H_{a}$ (real
magnetization is $4\pi M = B - H$, but $M_a$ is what can be
determined experimentally), in Fig.1(a), and the number of flux
quanta inside the sample, $N\Phi_o = \oint
(A+\frac{J_s}{|\Psi|^2})dl$, in Fig.1(b), as a function of 
$h=H_a/H_c(T)$.
\begin{figure}[tbh]
\begin{center}
\includegraphics[width=0.9\linewidth]{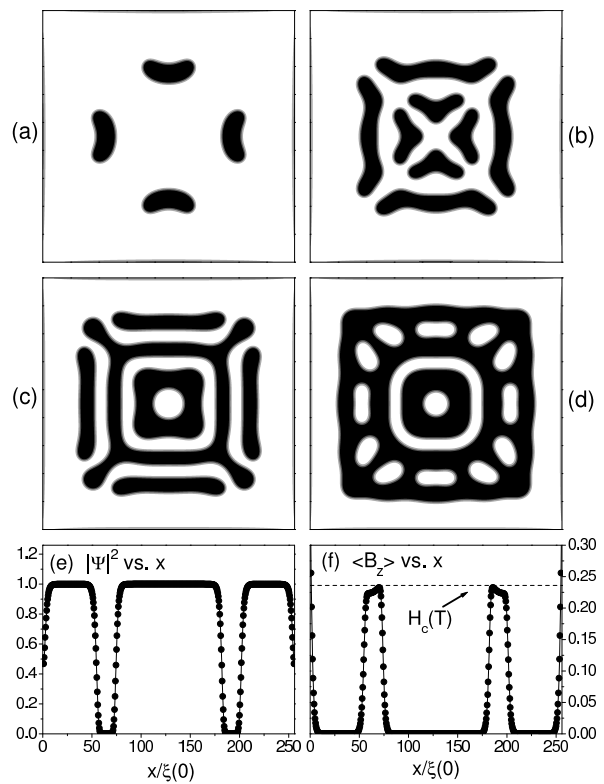}
\caption{(a)-(d) Spatial patterns of $|\Psi(r)|^2$ obtained continuously increasing 
$h=H_a/H_c$ from $h=0$. (a) h=0.58, (b) h=0.65, (c) h=0.72 and (d) h=0.78. 
Gray scale ranging from black for $|\Psi(r)|^2=0$ to white for
$|\Psi(r)|^2=1$. 
(e) and (f) show transversal cuts of $|\Psi(r)|^2$ and $\langle B_z \rangle$ taken at the center of one of the faces
for $h=0.58$ (Fig. 2(a)). 
Parameters: $\kappa=0.6$, $d=40\xi(0)$ and sample size $256\xi(0)\times256\xi(0)$. } 
\label{fig2}
\end{center}\end{figure}
\noindent

{\it (i) Slowly increasing the magnetic field}. We start from
$H_a=0$ with a state with $|\Psi({\bf R})|^2=1$ and 
$B({\bf R})=0$ and increase $H_a$ in
small steps, after reaching a stationary state for each field
(when $|\Delta E/E| <10^{-6}$; where $\Delta E$ is the change in energy
 between consecutive time steps). We observe a Meissner
state for $H_a < H_p = 0.56H_c$.
Surface barriers preclude the penetration of flux below $H_p$. 
The surface barrier in  macroscopic type-I
superconductors can lead to relatively large first penetration fields
$H_p$, which depend on the sample shape and
dimensions.\cite{fortini_barrier,castro} Here the smallness of
our system  strongly enhances this effect. We observe that at
$H\gtrsim H_p$ four long chunks of the normal phase, carrying
hundreds of flux quanta, enter from each side of the square and
equilibrate in the pattern shown in Fig.2(a). At a higher field
$H_{p,2}$, some other four chunks of flux enter and form the
pattern  seen in Fig.2(b). Further increasing the field, more
complex structures form, as shown in Fig.2(c) and (d).
The normal domains tend to stay in the centre of the sample,
leaving a flux-free zone near the edge.
Above $H_c$, flux has fully entered inside the 
system and there is only surface superconductivity 
until $H_{c3} > H_c$. The internal structure
of the domains is detailed in
Figs.2(e) and 2(f) which show transversal cuts of 
$|\Psi({\bf R})|^2$ and $B_z({\bf R})$ taken at the center of one of the faces for $h=0.58$. 
In Fig. 2(e) we see that in the normal regions there is a sharp drop to
zero of $|\Psi|^2$  and that  $B_z \sim H_c(T)$ (see Fig.2(f)).
The global structure of the patterns of the IS observed in
Figs.2(a)-(d)
follow the geometry of the sample and have the symmetry of the square. 
In general, we find that the entrance of
the normal phase occurs only for discrete values of penetration
fields $H_{p,i}$ where several flux quanta are nucleated at the
four sides of the square, while for $H_{p,i} < H_a < H_{p,i+1}$,
there is no flux entrance. This shows up as a saw-tooth behavior 
in the magnetization in Fig.1(a) and as a series of plateaus
and jumps in the number of flux quanta vs. $H_a$ in Fig.1(b). 
This type of behavior is similar to the results observed in mesoscopic
type-II systems \cite{geim} which also show a saw-tooth behavior
of the magnetization. However, while in
Ref.~\onlinecite{geim} each jump in $M_a$ corresponds to the
entrance of one quantum of flux (one vortex), here at each jump in
$M_a$ several hundreds of flux quanta have entered. 
Our results suggest the existence of a ``mesoscopic-like'' behavior in small type-I samples. 
This novel behavior appears when the linear size $L$ of the sample 
only allows for a small number of normal domains inside the system. 
This means that $L$  is not more than one order of magnitude larger than 
the periodicity of the patterns at intermediate fields. 
Indeed, we  have found similar  ``mesoscopic-like'' behavior 
for  sizes  in the range $L\le 256\xi(0)$.

\begin{figure}[tbh]
\begin{center}
\includegraphics[width=0.75\linewidth]{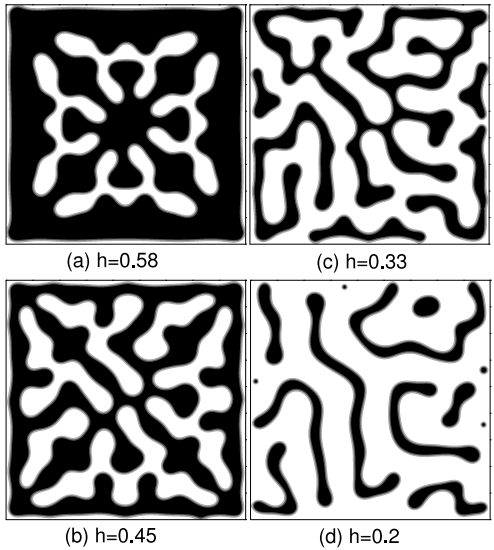}
\caption{Spatial patterns of $|\Psi(r)|^2 $ obtained decreasing
$h$ from the normal state at $h\gg 1$. Same parameters as in
Fig.2.} \label{fig3}
\end{center}\end{figure}
\noindent

{\it (ii) Slowly decreasing the magnetic field}. We start from
$H_a > H_c$ with a state with $\Psi=0$ and $B=H_a$ and decrease
$H_a$ in small steps, after reaching a stationary state for each
field. The resulting intermediate state patterns are shown in
Fig.3. When $H_a \approx H_c$ the superconducting phase enters
into the sample and the total number of flux quanta is reduced. At
first, the superconducting phase forms four chunks embedded within
the normal phase, which nearly follow the square symmetry of the
system, as shown in Fig.3(a). However, we observe that when
decreasing the field the square symmetry is always broken in the
patterns. 
The breaking of symmetry is stronger
the further we decrease the field. In this way, labyrinthine
patterns are formed at mid-range fields, as can be seen in
Figs.3(b) and (c). 
In general,
the expulsion of flux
occurs gradually when decreasing $H_a$ as shown in Fig.1(b). 
For low fields we see that thin stripes of normal phase are trapped
within the sample, as shown in Fig.3(d). 
The difficulty for expelling flux is due to the
surface barrier and results in a positive magnetization as a
function of $h$ as shown in Fig.1(a).  Even at $h=0$,  a small 
amount of flux remain trapped in the sample, as evidenced 
in Fig.1(b), where the number of flux quanta is
finite at $h=0$, and in Fig.1(a), where $M_a > 0$ at $h=0$. 
In experiments in macroscopic samples  it has been
observed that some trapped flux remains in the system at $h=0$ 
when decreasing the field \cite{Huebener72} and
in some Sn films a 
positive magnetization has been obtained when decreasing $H_a$.\cite{miller_cody}
It is interesting to mention that a
similar positive magnetization was observed in mesoscopic type-II
superconductors \cite{geim} when decreasing $H_a$, and attributed to
the importance of surface barriers.

\begin{figure}[tbh]
\begin{center}
\includegraphics[width=0.75\linewidth]{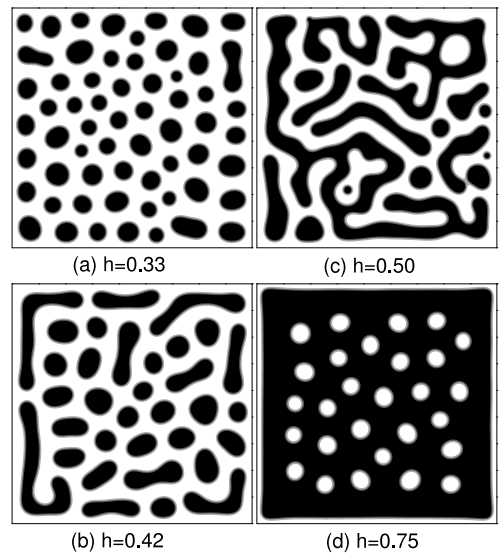}
\caption{$|\Psi(r)|^2$ patterns obtained using random initial
conditions. Same parameters as in Fig.2.} 
\label{fig4}
\end{center}\end{figure}
\noindent

{\it (iii) Random initial conditions}. The structures of the
IS discussed above, obtained either increasing or
decreasing the magnetic field, are strongly influenced by the
surface barriers and/or the geometry of the small sample 
simulated. In a film of large linear size $L$ the demagnetization
factor $N$ is such that $ 1-N\propto d/L \approx 0$, and therefore we expect that
$B\approx H_a$. To obtain stationary states more typical of the
bulk behavior of large samples, we start with a initial condition
with random values of ${\bf A}$ and $\Psi$, such that we satisfy
$\langle B \rangle =H_a$  from the start, and that the initial
state is superconducting in average, $\langle |\Psi|^2\rangle > 0$. 
We performed simulations with this initial condition for
different values of $h$, and let the system to evolve for each case,
using a stronger criterion for assuming stationarity: 
$|\Delta E/E| <10^{-9}$. 
We obtain that in the stationary state most of the flux 
remains inside the sample and $\langle B_z\rangle-H_a\approx 0$ 
as can be seen in Fig.1(a). The structures obtained are shown in Fig.~4. For
low fields, we observe in Fig.4(a) that the intermediate
state consists of almost circular droplets of the normal phase.
For higher fields, the droplets start to coalesce into long
lamellar-like domains, as seen in Fig.4(b). At intermediate
fields, as shown in Figs.4(c), labyrinthine patterns of
stripes of the normal phase are formed. For high fields close to
$H_c$ we observe almost circular droplets of the superconducting
phase embedded within the mostly normal phase, see Fig.4(d).
Similar type of structures, with droplets of one or the other
phase at low and high fields, and with labyrinthine patterns of
stripes at mid range fields has been observed experimentally 
for example in Figs.2.8(a)-(f) of Ref.\onlinecite{Huebener} for a lead thin film.
An important feature we find in our small system is that there
is a thin layer of superconducting phase at the surface 
(see Fig.~4), which allows for the screening Meissner currents.

\begin{figure}[tbh]
\begin{center}
\includegraphics[width=0.75\linewidth]{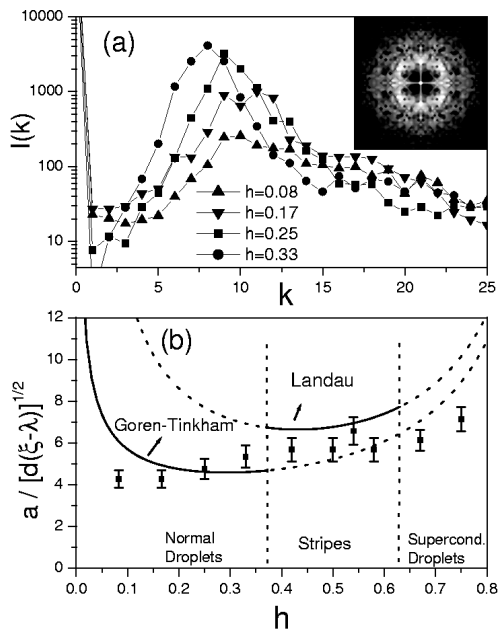}
\caption{ (a) Spectral intensities obtained from $|\Psi(r)|^2$ at
different magnetic fields, a maximum at $k_0$ is observed. The
inset shows the Fourier transform of Fig. 4(a). (b) Periodicity
$a=2\pi/k_0$ of each structure as a function of $h$. The lines
correspond to the Landau and the Goren-Tinkham models.}
\label{fig5}
\end{center}\end{figure}
\noindent

We analyze the structures obtained in Fig.4, by calculating the
spectral transform of the superconducting order parameter,
 $I({\bf k})=\left|\int d{\bf r} |\Psi({\bf r})|^2 \exp(i {\bf
k}\cdot{\bf r})\right|^2$, which is shown in Fig.5(a). 
The non-periodicity and complex structure of the patterns
results in very broad maxima in $I({\bf k})$
at finite wave vectors $k_0 = 2\pi/a$ 
which define a typical length scale $a$.
In the case of low and large fields, $a$ would 
correspond to the typical distance
between droplets, while for mid-range fields, $a$ would correspond to the
average widths of the stripes in the labyrinthine patterns. 
We plot $a$ in Fig.5(b) and compare it with the Landau model of stripes \cite{Landau} and
with a model of  Goren and Tinkham for a periodic array of droplets
or ``flux spots''.\cite{Goren_Tinkham} 
We see that the Landau model agrees qualitatively  with
the results obtained at mid-range $h$
(for these fields the patterns of Fig.4 can have a mixture
of ``stripes'' with a few droplets, which  make $a$ smaller
than the Landau value).
On the other hand, at low $h$ the Goren-Tinkham
model does not agree well  with the  size of the droplets obtained. 
Also in some experiments \cite{kirchner,jeudy} 
it has been found a departure from the
Goren-Tinkham model at low fields.

In conclusion, our simulations predict that the strong influence of 
the surface barriers in small type-I samples will lead to a saw-tooth 
behavior of the magnetization and very geometric patterns when
slowly increasing $H_a$, and to a positive magnetization and 
symmetry-breaking structures when slowly decreasing $H_a$.
These results suggest the existence of a ``mesoscopic-like'' behavior in the IS 
when the sample linear size is of the 
order of a few times the periodicity of the patterns. 
It will be interesting if experiments on small samples of type-I 
superconductors could be performed. 

We acknowledge discussions with E. Jagla, M. Menghini, F. de la
Cruz and financial support from ANPCYT, CNEA and
Conicet. 
ADH acknowledges support from CLAF and Fundaci\'on Antorchas.


\begin{thebibliography}{}

\bibitem{Landau} L.D. Landau,
Zh. Eskp. Teor. Fiz. {\bf 7}, 371 (1937).

\bibitem{Huebener} R.P. Huebener, {\it Magnetic Flux Structures in
Superconductors} (Springer-Verlag, New York, 1979).


\bibitem{Huebener72}
R.~P. Huebener, R.~T. Kampwirth, y V.~A. Rowe, Cryogenics {\bf 12},  100
  (1972).


\bibitem{kirchner}
A. Kiendl and H. Kirchner, J. Low Temp. Phys., {\bf 14}, 349 (1974).

\bibitem{miller_cody} R. E. Miller and G. D. Cody, Phys. Rev. {\bf
173}, 494 (1968).


\bibitem{seul_andelman}
M. Seul and D. Andelman, Science {\bf 267}, 476 (1995).


\bibitem{castro} 
H. Castro, B. B. Dutoit, A. Jacquier, M. Baharami, and L. Rinderer, 
Phys. Rev. B {\bf 59}, 596 (1999).

\bibitem{reisin_lipson} 
C. R. Reisin and S. G. Lipson, Phys. Rev. B {\bf 61}, 4251 (2000).

\bibitem{egorov} 
V. S. Egorov, G. Solt, C. Baines, D. Herlach, and U. Zimmermann, 
Phys. Rev. B {\bf 64}, 024524 (2001).

\bibitem{jeudy} 
V. Jeudy, C. Gourdon, and T. Okada, Phys. Rev. Lett. {\bf 92}, 147001 (2004).

\bibitem{mariela} M. Menghini and R. J. Wijngaarden, preprint.


\bibitem{maki_lasher_callaway}
K. Maki, Ann. Phys. {\bf 34}, 363 (1965); G. Lasher,  Phys. Rev.
{\bf 154}, 345 (1967); D.J.E. Callaway, Ann. Phys. {\bf 213}, 166
(1992).

\bibitem{hubert} 
A. Hubert, Phys. Stat. Sol. {\bf 24}, 669 (1967).

\bibitem{Goren_Tinkham} 
R.N. Goren and M. Tinkham, J. Low Temp. Phys. {\bf 5}, 465 (1971).

\bibitem{fortini_barrier} 
E. Fortini and E. Paumier, Phys. Rev B {\bf 14}, 55 (1976).

\bibitem{SL} 
J. M. Simonin and A. L\'{o}pez, J. Low Temp. Phys. {\bf 41}, 105 (1980).

\bibitem{Dorsey91} 
H. Frahm, S. Ullah, and A. T. Dorsey, Phys. Rev. Lett. {\bf 66}, 3067 (1991); 
F. Liu, M. Mondello, and N. Goldenfeld, {\it ibid.} {\bf 66}, 3071 (1991).


\bibitem{dorsey96} 
R. E. Goldstein, D. P. Jackson, and A. T. Dorsey, Phys. Rev. Lett. {\bf 76}, 3818 (1996); 
A.T. Dorsey and R.E. Goldstein, Phys. Rev. B {\bf 57}, 3058 (1998).

\bibitem{narayan} 
H. Bokil and O. Narayan, Phys. Rev. B {\bf 56}, 11195
(1997); O. Narayan, Phys. Rev. Lett. {\bf 81}, 5035
(1998); R. E. Goldstein and A. T. Dorsey, {\it ibid.} {\bf 81}, 5036
(1998).

\bibitem{choksi}  R. Choksi, R.V.
Kohn, and F. Otto, J. Nonlinear Sci. {\bf 14}, 119 (2004).


\bibitem{geim}
A.~K. Geim, I.~V. Grigorieva, S.~V. Dubonos, J. Lok, J. Maan, A.
Filipov, and F. Peeters, Nature {\bf 390},  259  (1997).


\bibitem{brandt} 
E. H. Brandt, Phys. Rev. Lett. {\bf 74}, 3025 (1995).

\bibitem{dj96} 
D. Dom\'{\i}nguez and J. V. Jos\'e, Phys. Rev. B {\bf 53}, 11692 (1996).


\end{thebibliography}
\end{document}